\documentclass[english]{paper}
\usepackage{lmodern}
\usepackage{lmodern}
\usepackage[T1]{fontenc}
\usepackage[latin9]{inputenc}
\usepackage{geometry}
\usepackage{color}
\usepackage{babel}
\usepackage{float}
\usepackage{amsmath}
\usepackage{amssymb}
\usepackage{graphicx}
\usepackage[unicode=true,pdfusetitle,
 bookmarks=true,bookmarksnumbered=false,bookmarksopen=false,
 breaklinks=false,pdfborder={0 0 1},backref=false,colorlinks=false]
 {hyperref}
\hypersetup{
 colorlinks,linkcolor=blue,citecolor=blue,urlcolor=blue}

\makeatletter

\newcommand*\LyXThinSpace{\,\hspace{0pt}}
\providecommand{\tabularnewline}{\\}

\newcommand{\lyxaddress}[1]{
	\par {\raggedright #1
	\vspace{1.4em}
	\noindent\par}
}

\@ifundefined{date}{}{\date{}}
\usepackage{babel}

\usepackage{blindtext}

\usepackage{dsfont}\usepackage{bbm}\IfFileExists{lmodern.sty}{\usepackage{lmodern}}{}
\usepackage{babel}

\makeatother

\begin{document}

\title{Microscopic linear response theory of spin relaxation and relativistic
transport phenomena in graphene }

\author{Manuel Offidani\,$^{*}$, Roberto Raimondi\,$^{\dagger}$ and Aires
Ferreira\,$^{*}$}
\maketitle

\lyxaddress{$^{*}$Department of Physics, University of York, York YO10 5DD,
United Kingdom}

\lyxaddress{$^{\dagger}$Dipartimento di Matematica e Fisica, Università Roma
Tre, 00146 Rome, Italy}
\begin{abstract}
We present a unified theoretical framework for the study of spin dynamics
and relativistic transport phenomena in disordered two-dimensional
Dirac systems with pseudospin-{}-spin coupling. The formalism is applied
to the paradigmatic case of graphene with uniform Bychkov-{}-Rashba
interaction and shown to capture spin relaxation processes and associated
charge-to-spin interconversion phenomena in response to generic external
perturbations, including spin density fluctuations and electric fields.
A controlled diagrammatic evaluation of the generalized spin susceptibility
in the diffusive regime of weak spin-orbit interaction allows us to
show that the spin and momentum lifetimes satisfy the standard Dyakonov-Perel
relation for both weak (Gaussian) and resonant (unitary) nonmagnetic
disorder. Finally, we demonstrate that the spin relaxation rate can
be derived in the zero-frequency limit by exploiting the SU(2) covariant
conservation laws for the spin observables. Our results set the stage
for a fully quantum-{}-mechanical description of spin relaxation in
both pristine graphene samples with weak spin-{}-orbit fields and
in graphene heterostructures with enhanced spin-{}-orbital effects
currently attracting much attention. 
\end{abstract}
\newpage{}

\section{Introduction\label{sec:Introduction}}

\subsection{Spin Relaxation in Graphene}

Graphene is considered a promising material for spintronics applications
due to its negligible hyperfine interactions and low spin--orbit
coupling (SOC) \cite{Huertas_PRB2006,Konschuh_10}. Early theoretical
estimates hinted at ultra-long spin lifetime ($\tau_{s}\approx$1--100\,$\mu$s)
\cite{Pesin_NatMat2012}, whereas experiments found $\tau_{s}$ to
be limited to a few nanoseconds \cite{Han_NatNano2014}. The microscopic
mechanisms responsible for the relatively fast spin relaxation in
high-mobility graphene samples remain controversial \cite{Roche_Valenzuela_14},
but recent findings indicate that spinful scatterers, such as magnetic
adatoms, are the primary cause of spin relaxation \cite{m_Avila_15,m_Lundeberg_13,m_Omar_17,m_Raes_16}.

The spin dynamics in graphene is conventionally probed by means of
nonlocal transport measurements \cite{Silsbee_85,Jedema_01}. In this
approach, a spin current is injected from ferromagnetic electrodes
into the graphene channel and allowed to diffuse under the effect
of a perpendicular magnetic field. The Larmor precession of the electron's
spin about the external field modulates the average spin accumulation
detected away from the injection point (Hanle curve), resulting in
a bona fide spin signal from which $\tau_{s}$ can be deduced. Such
Hanle precession measurements found a large spread in $\tau_{s}$
from tens of picoseconds up to a few nanoseconds \cite{Tombros_Nature2007,Josza_PRB2009,Popinciuc_PRB2009,Han_PRL2010,Yang_PRL2011,Han_PRL2011,Pi_PRL2011,Jo_PRB2011,Zomer_PRB2012,Drogler_NanoLett2014,Drogler_NanoLett2016},
reflecting the different sample preparation and device fabrication
methods. Theoretical studies have revealed a number of possible spin
relaxation sources, including magnetic impurities, spin--orbit active
adatoms, ripples and other substrate effects \cite{Guinea_09,Huertas_PRL2009,Ertler_PRB2009,Kochan_PRL2014,Maasen_PRB2011,Volmer_PRB2013,Federov13,Soriano_2DMat2015}.
Numerical approaches have provided further insight into the relaxation
mechanisms, enriching the scenario to include the impact of electron-hole
puddles, pseudospin-spin coherence and ballistic effects \cite{Tuan_14,Tuan_Sci2016,Cummings_PRL2016}.
Despite the relatively short $\tau_{s}$ in of clean samples, the
high charge carrier mobility allows spins to diffuse over extremely
long distances up to 13\,$\mu$m at room temperature \cite{Wojtaszek_13-1,Guimaraes12-1,Ayn=00003D0000E9s_15-1}. 

The paradigmatic model for studies of spin relaxation in graphene
is the two-dimensional (2D) Hamiltonian of massless Dirac fermions
supplemented with a (uniform or random) Bychkov-Rashba interaction
\cite{Bychkov_Rashba_84}. This type of SOC has its origin in perturbations
breaking the inversion symmetry, which include substrate-induced electric
fields, adatoms, and ripple-induced gauge fields \cite{Pesin_NatMat2012,Han_NatNano2014}.
The Bychkov-Rashba interaction in graphene (hereafter referred to
as Rashba SOC) can be seen as a non-Abelian gauge field that couples
to the intrinsic pseudospin of Dirac fermions, enabling spin relaxation
upon impurity scattering \emph{e.g.,} via the familiar Dyakanov-Perel
(DP) mechanism \cite{Wu10}. 

Graphene with random Rashba SOC has been recently shown to host novel
charge-to-spin conversion effects by means of a quantum extension
of the Boltzmann transport theory \cite{Huang_PRB2016,Huang_PRL2017}.
Previous theoretical descriptions of spin relaxation in such 2D Dirac-Rashba
models were based on semiclassical approximations \cite{Ochoa_12,Zhang_NJP2012}.
On the other hand, a \textcolor{black}{fully quantum-mechanical }theory
of spin--orbit-coupled transport\textcolor{black}{{} in the static
(DC) limit has been formulated recently by the authors \cite{Milletari_PRL2017,Offidani_PRL2017}.
Analogously to the 2D electron gas (2DEG) case \cite{Dimitrova_PRB2005,Raimondi_PRB2006,Raimondi_Annalen2012},
it was shown that impurity scattering corrections exactly balance
the intrinsic generation of spin Hall current for spin-independent
disorder, $\langle\mathcal{J}_{\text{SH}}\rangle_{\boldsymbol{\mathcal{E}}}=0$,
where $\boldsymbol{\mathcal{E}}$ is an external DC electric field
\cite{Milletari_PRL2017}. The vanishing of the spin Hall effect in
this model is connected to the establishment of a robust nonequilibrium
in-plane spin polarization $\langle\text{\textbf{S}}\rangle_{\boldsymbol{\mathcal{E}}}\neq0$
with $\text{\textbf{S}\ensuremath{\perp}}\boldsymbol{\mathcal{E}}$,
known as inverse spin-Galvanic effect (ISGE) \cite{Offidani_PRL2017}.
However, a time-dependent framework able to unveil how the steady
state is reached within the 2D Dirac-Rashba mod}el is yet to be developed.
In this paper, we address this problem. We derive the coupled spin-charge
drift-diffusion equations for nonmagnetic disorder and generic homogeneous
perturbations by means of the diagrammatic technique for disordered
electrons. A similar approach has been adopted very recently in the
context of 2DEGs with both Bychkov-Rashba and Dresselhaus interactions
\cite{Maleki_CondMatt2017}, where it was shown perfect agreement
between the Kubo diagrammatic formalism and the Keldysh SU(2) gauge
theory \cite{Gorini_PRB2017}. In this work, we extend the standard
quantum diagrammatic formalism to accommodate the enlarged $2$ (spin)
$\otimes$ 2 (pseudospin) Clifford structure of the 2D Dirac--Rashba
model leading to a 16-dimensional diffuson operator in the absence
of intervalley scattering. We find that the typical DP relation connecting
the spin relaxation time (SRT) and the momentum lifetime in the weak
SOC regime, that is $\tau_{s}\propto\tau^{-1}$ for $\lambda\tau\ll1$,
where $\lambda$ is the SOC strength, holds at all orders in the scattering
potential strength. The meaning and interpretation of our results
for the SRTs can be also clarified by the SU(2) covariant conservation
laws inherent to the diagrammatic (perturbative) structure, whose
usage allows us derive the DP relation even in the zero-frequency
limit. In particular, we provide the analytical expression of $\tau_{s}$
in the unitary limit of very strong potential scattering.

\subsection{Dirac--Rashba model\label{subsec:Dirac=00003D002013Rashba-model}}

The effective low-energy Hamiltonian describing the electronic properties
of 2D Dirac fermions subject to a uniform Rashba interaction around
the $K$ point reads as \cite{Note_basis}

\begin{equation}
H=\int d\mathbf{x}\,\,\Psi^{\dagger}(\mathbf{x})\,\left[v\,\boldsymbol{\sigma}\cdot\mathbf{p}+\lambda\,\left(\boldsymbol{\sigma}\times\mathbf{s}\right)\cdot\hat{z}+V(\mathbf{x})\right]\Psi(\mathbf{x})\,\,,\label{eq:Hamiltonian}
\end{equation}
where $v$ is the bare velocity of massless Dirac fermions, $\mathbf{p}=-\imath\boldsymbol{\nabla}$
is the 2D kinematic momentum operator, $\lambda$ is the SOC strength
and $\sigma_{i},s_{i}$ ($i=x,y,z$) are Pauli matrices associated
with sublattice (pseudospin) and spin degrees of freedom, respectively.
Here, $V(\text{\textbf{x}})$ is a disorder potential describing elastic
scattering from nonmagnetic short-range impurities. For simplicity,
in this work we neglect intervalley scattering processes, which in
the pure Rashba model can renormalize the momentum lifetime but are
not expected to impact fundamentally the spin dynamics \cite{Milletari_PRL2017}.
Thus it suffices to consider the low-energy dynamics around the $K$
point.

\begin{figure}
\centering{}\includegraphics[width=0.6\columnwidth]{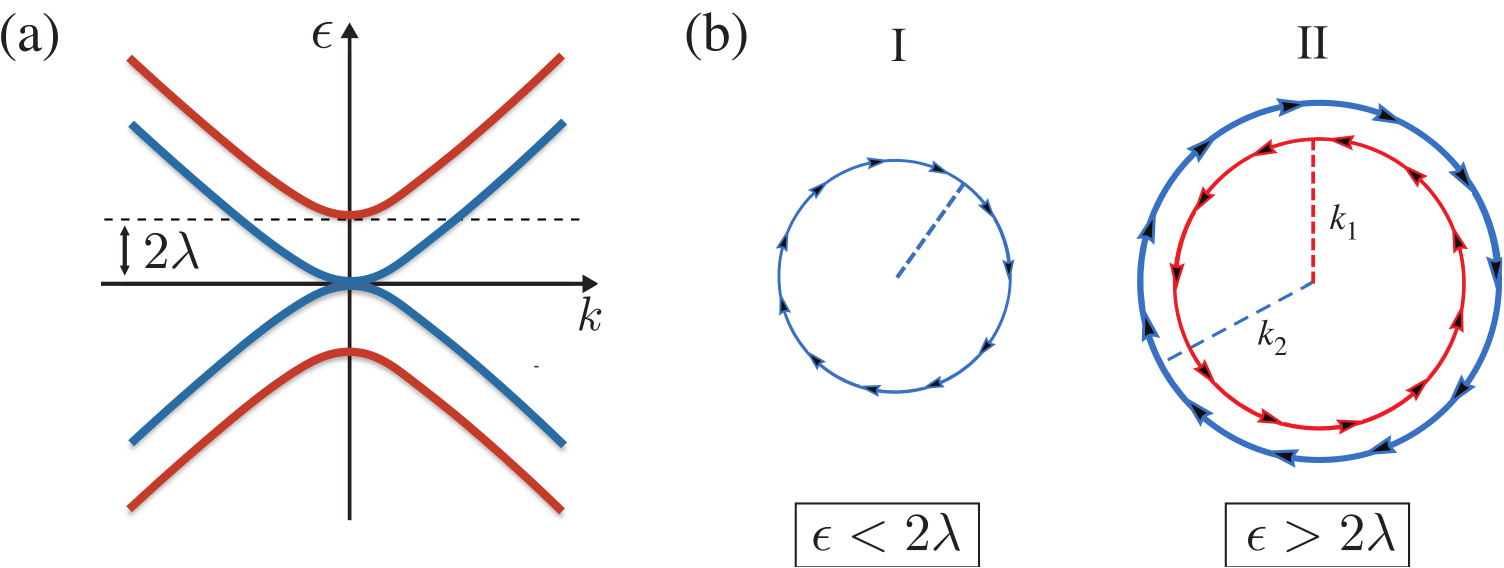}\caption{\label{fig:01-band_structure}(a) Energy dispersion around the $K$
point. The splitting of the Dirac bands leads to a spin gap or \emph{pseudogap}.
(b) Tangential winding of the spin texture in regimes I and II.}
\end{figure}

The energy dispersion relation of the free Hamiltonian $H_{0}=H-V$
in Eq.\,(\ref{eq:Hamiltonian}) is 
\begin{equation}
\epsilon_{\mu\nu}(\mathbf{k})=\mu\lambda+\nu\sqrt{\lambda^{2}+v^{2}|\mathbf{k}|^{2}}\,,\label{eq:spectrum}
\end{equation}
where $\mu,\nu=\pm1$ labels the various subbands (Fig.\,\ref{fig:01-band_structure}(a)).

The Rashba interaction aligns the electron spin at right angles to
the wavevector, the so-called spin--momentum locking configuration
(Fig.\,\ref{fig:01-band_structure}(b)) \cite{Schwab_EPL2011,Hsieh_TIs_09}.
For Fermi energy $|\epsilon|>2|\lambda|$ (region II), the split Fermi
surface displays counter-rotating spin textures reminiscent of (nonchiral)
2DEGs with Rashba interaction \cite{Bychkov_Rashba_84}. A regime
(pseudogap, region I) where the Fermi energy intersects a single subband,
with electronic states having well-defined spin helicity, extends
for energies $|\epsilon|<2|\lambda|$. In the conventional 2DEG this
circumstance only happens at a single point \emph{i.e.}, the intersection
between the parabolic bands \cite{Brosco_PRL2016}. Importantly, the
spin texture of energy bands in the 2D Dirac--Rashba model is modulated
by the band velocity \emph{i.e.}, 
\begin{equation}
\langle\mathbf{s}\rangle_{\mu\nu\mathbf{k}}=-\mu\langle\boldsymbol{\sigma}\rangle_{\mu\nu\mathbf{k}}\times\mathbf{\hat{z}}\,,\label{eq:texture}
\end{equation}
where $\langle\boldsymbol{\sigma}\rangle_{\mu\nu\mathbf{k}}=(1/v)\nabla_{\mathbf{k}}\epsilon_{\mu\nu}(\mathbf{k})$
is the pseudospin polarization vector. The entanglement between pseudospin
and spin degrees of freedom in the model is responsible for a rich
energy dependence of transport coefficients \cite{Milletari_PRL2017,Offidani_PRL2017}.
For brevity of notation, we assume $\epsilon,\lambda>0$ in the remainder
of the work.

\subsection{Disorder effects}

The random potential in Eq.~(\ref{eq:Hamiltonian}) affects the spin
dynamics by inducing elastic transitions between electronic states
$(\mu\nu\mathbf{k})\rightarrow(\mu^{\prime}\nu^{\prime}\mathbf{k}^{\prime})$
associated with different effective Larmor fields, $\boldsymbol{\Omega}_{\mu\nu\mathbf{k}}=\lambda\langle\mathbf{s}\rangle_{\mu\nu\mathbf{k}}\approx-\mu\nu\lambda\,\hat{\mathbf{k}}\times\hat{z}$
for $\epsilon\gg\lambda$. This random change in the spin precession
axis is responsible for the irreversible loss of spin information.
To describe the effects of disorder, we employ standard many-body
perturbation theory methods. We work within the zero-temperature Green's
function formalism.

The retarded(R)/advanced(A) single-particle Green's function ($a=A,R\equiv-,+$)
is

\begin{equation}
G^{a}(\mathbf{x},\mathbf{x}^{\prime};t-t^{\prime})=\mp\imath\left\langle 0|\mathcal{T}\left[\Psi(\mathbf{x},t),\Psi^{\dagger}(\mathbf{x}^{\prime},t^{\prime})\right]|0\right\rangle \theta(\pm t\mp t^{\prime}),\label{eq:time-ordered-G}
\end{equation}
where $\mathcal{T}$ is the time-ordering symbol and $\theta(.)$
is the Heaviside step function. Changing to the energy domain, one
obtains 
\begin{equation}
G^{a}(\mathbf{x},\mathbf{x}^{\prime};\epsilon)=\langle\mathbf{x}^{\prime}|\frac{1}{[G_{0}^{a}(\epsilon)]^{-1}-V}|\mathbf{x}\rangle\,,\label{eq:GR_GA}
\end{equation}
where $G_{0}^{a}(\epsilon)=(\epsilon+\imath v\boldsymbol{\sigma}\cdot\mathbf{\boldsymbol{\nabla}}+\lambda\,\left(\boldsymbol{\sigma}\times\mathbf{s}\right)\cdot\hat{z}\pm\imath0^{+})^{-1}$
is the Green's function of free 2D Dirac--Rashba fermions.

\begin{figure}
\centering{}\includegraphics[width=0.9\textwidth]{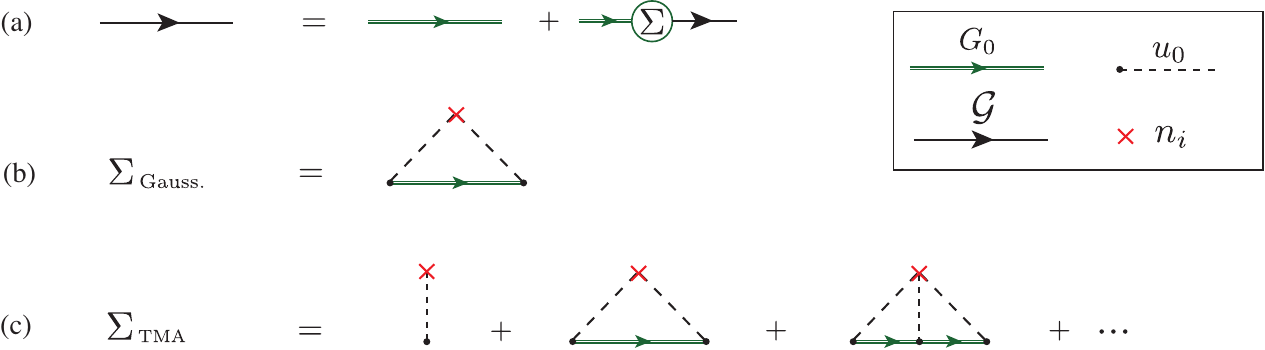}\caption{\label{fig:02_Diagrams_SelfEnergy}(a) Dyson equation for the disordered
averaged Green's function. (b-c) Approximation schemes for evaluation
of the self energy: Gaussian (b) and $T$-matrix approximation (TMA)
(c). Box shows Feynman rules for the disorder potential insertions
(dashed lines) and impurity density insertion (red crosses).}
\end{figure}

The central quantity in our approach is the disorder averaged Green's
function, $\mathcal{G}^{a}(\mathbf{x}-\mathbf{x}^{\prime},\epsilon)=\overline{G^{a}(\mathbf{x},\mathbf{x}^{\prime};\epsilon)}$,
where the bar $\overline{\cdot}$ denotes the average over all impurity
configurations (Fig.~\ref{fig:02_Diagrams_SelfEnergy}(a)). Its momentum
representation is 
\begin{equation}
\mathcal{G}_{\mathbf{k}}^{a}(\epsilon)=\frac{1}{[\mathcal{G}_{0\mathbf{k}}^{a}(\epsilon)]^{-1}-\Sigma_{\mathbf{k}}^{a}(\epsilon)}\,,\label{eq:av_Greens_Function_}
\end{equation}
where $\mathcal{G}_{0\mathbf{k}}^{a}(\epsilon)$ is the Fourier transform
of $G_{0}^{a}(\mathbf{x}-\mathbf{x}^{\prime};\epsilon)$ and 
\begin{equation}
\Sigma_{\mathbf{k}}^{a}(\epsilon)=\int d\mathbf{(x-x^{\prime})}\,e^{-\imath\mathbf{k}(\mathbf{x}-\mathbf{x}^{\prime})}\overline{\langle\mathbf{x}^{\prime}|V\frac{1}{1-G_{0}^{a}(\epsilon)V}|\mathbf{x}\rangle}\,,\label{eq:self-energy-1}
\end{equation}
is the disordered averaged self energy within the noncrossing approximation.
The latter neglects coherent multiple impurity scattering corrections,
which is justified in the diffusive regime with $\epsilon\tau\gg1$
\cite{Milletari_Ferreira_QSHE_QTheory_16}. The self-energy induced
by short-range impurities is $\mathbf{k}$-independent, $\Sigma_{\mathbf{k}}^{a}(\epsilon)\equiv\Sigma^{a}(\epsilon)$,
and hence we drop this index in what follows.

To account for the characteristic resonant (unitary) scattering regime
of graphene with relaxation time $\tau\propto\epsilon$ \cite{Ferreira_11_res,Ferreira_PRL2014},
we adopt a $T$-matrix approach by evaluating the self energy $\Sigma^{a}(\epsilon)$
at all orders in $V$. We obtain 
\begin{equation}
\Sigma^{a}(\epsilon)=n_{i}\frac{u_{0}}{1-u_{0}g_{0}^{a}(\epsilon)}+O(n_{i}^{2})=n_{i}T^{a}(\epsilon)\,,\label{eq:TMatrix_SelfEnergy}
\end{equation}
where $u_{0}$ parameterizes the scattering strength of the spin-transparent
(scalar) impurities, $n_{i}$ is the impurity areal density and $T^{a}(\epsilon)$
is the single-impurity $T$-matrix. Note that multiple impurity scattering
diagrams $\propto O(n_{i}^{2})$ can be neglected in the limit $\epsilon\tau\gg1$
i.e., away from the Dirac point (refer to Sec.~\ref{subsec:Discussion}
for a brief discussion of the spin relaxation within the full noncrossing
approximation). We have also introduced 
\begin{equation}
g_{0}^{a}(\epsilon)=\,g_{0,0}^{a}(\epsilon)\gamma_{0}+g_{0,zz}^{a}(\epsilon)\,\gamma_{zz}+g_{0,r}^{a}(\epsilon)\,\gamma_{r}\,,\label{eq:integrated_clean_Green_function}
\end{equation}
as the momentum integrated Green's function of the clean system {[}cf.
Eq.\,(\ref{eq:G_0_explicit}) of Appendix A{]}, where $\gamma_{0}\equiv\sigma_{0}s_{0}$
is the $4\times4$ identity matrix, $\gamma_{zz}=\sigma_{z}s_{z}$,
$\gamma_{r}=\left(\boldsymbol{\sigma}\times\text{\textbf{s}}\right)_{z}$
and 
\begin{align}
g_{0,0}^{a}(\epsilon) & =-\frac{1}{8\pi v^{2}}\left[\epsilon\left(\mathcal{L}_{\textrm{II}}(\epsilon)+a\imath\pi\,\theta_{\text{II}}(\epsilon)\right)+\lambda\left(\mathcal{L}_{\textrm{I}}(\epsilon)+a\imath\pi\,\theta_{\text{I}}(\epsilon)\right)\right]\,,\label{eq:g00}\\
g_{0,zz}^{a}(\epsilon) & =-\frac{\lambda}{8\pi v^{2}}\left(\mathcal{L}_{\textrm{I}}(\epsilon)+a\imath\pi\,\theta_{\text{I}}(\epsilon)\right)\,,\label{eq:gzz}\\
g_{0,r}^{a}(\epsilon) & =+\frac{\epsilon}{16\pi v^{2}}\left(\mathcal{L}_{\textrm{I}}(\epsilon)+a\imath\pi\,\theta_{\text{I}}(\epsilon)\right)\,.\label{eq:gr}
\end{align}
In the above, $\theta_{\textrm{I}(\textrm{II})}(\epsilon)=\theta(\epsilon+2\lambda)\mp\theta(\epsilon-2\lambda)$
selects the energy regime and 
\begin{equation}
\mathcal{L}_{\textrm{I}(\textrm{II})}(\epsilon)=\log\left|\frac{\Lambda^{2}}{\epsilon(\epsilon+2\lambda)}\right|\mp\log\left|\frac{\Lambda^{2}}{\epsilon(\epsilon-2\lambda)}\right|\,,\label{eq:LI_II}
\end{equation}
with $\Lambda$ denoting the ultraviolet cutoff of the low-energy
theory \cite{Ferreira_11_res}.

The self energy simplifies in two important limiting cases: (i) weak
Gaussian disorder ($|u_{0}|\ll|g_{0}^{a}|^{-1}$) and (ii) unitary
disorder ($u_{0}\rightarrow\pm\infty$). In the weak scattering regime,
it suffices to only take into account the 'rainbow' diagram with two
impurity lines in the Dyson expansion; see Fig.\,\ref{fig:02_Diagrams_SelfEnergy}\,(b).
For scalar disorder this approximation is equivalent to assuming that
the= disorder potential satisfies white-noise statistics \cite{Milletari_Ferreira_QSHE_QTheory_16}
\begin{align}
\langle V(\mathbf{x})\rangle & =0\,,\label{eq:white_noite_1}\\
\langle V(\mathbf{x})V(\mathbf{x}^{\prime})\rangle & =n_{i}u_{0}^{2}\,\delta(\mathbf{x}-\mathbf{x}^{\prime})\,.\label{eq:white_noite_2}
\end{align}
In this case we have 
\begin{equation}
\left.\Sigma^{a}(\epsilon)\right|_{\textrm{Gauss.}}=n_{i}u_{0}^{2}\,g_{0}^{a}(\epsilon)\,.\label{eq:SelfEnergyBA}
\end{equation}
The real part of the self-energy provides a parametrically small renormalization
of the band structure, which can be safely neglected in the diffusive
regime of interest \cite{Milletari_PRL2017}. We thus find 
\begin{align}
\Sigma^{R/A} & =\mp\imath\,n_{i}(\eta_{0}\gamma_{0}+\eta_{r}\,\gamma_{r}+\eta_{zz}\,\gamma_{zz})\,,\label{eq:self_energy_final}
\end{align}
where the functions $\eta_{0},\eta_{r},\eta_{zz}$, proportional to
the imaginary parts of Eqs.\,(\ref{eq:g00})-(\ref{eq:gr}), have
different forms depending on the Fermi level position. In this work,
we will restrict the analysis to diffusive systems with weak SOC $\lambda\tau\ll1$
and $\epsilon\gg\lambda$. It is thus convenient to express the various
quantities in $\Sigma^{a}(\epsilon)$ in terms of the quasiparticle
broadening in regime II, i.e., 
\begin{equation}
\frac{1}{2\tau}\equiv n_{i}\eta_{0}|_{\epsilon>2\lambda}.\label{eq:def_tau_0}
\end{equation}
Explicitly, we have 
\begin{align}
\left.\frac{1}{2\tau}\right|_{\textrm{Gauss.}}=n_{i}\,\frac{u_{0}^{2}\epsilon}{4v^{2}},\,\,\eta_{zz}=0,\,\,\eta_{r}=0.\label{eq:def_plus}
\end{align}
For a typical choice of parameters, say, $n_{i}=10^{12}$ cm$^{-2}$,
$u_{0}=1$ eV ($u_{0}$ is in units of eV$\cdot$nm$^{-2}$) and $\epsilon=50$
meV, one finds $\left.\tau\right|_{\text{Gauss}}\simeq1.14$ ps, which
is representative of clean graphene samples \cite{Ferreira_11_res}. 

Within the $T$-matrix formalism, the nondiagonal part of $\Im\Sigma^{a}(\epsilon)$
acquires a finite value. However, in the unitary limit of strong potential
scattering ($u_{0}\rightarrow\infty$), we have $\Sigma^{a}(\epsilon)=-n_{i}/g_{0,0}^{a}(\epsilon)$
and we recover a scalar self-energy, with 
\begin{align}
 & \left.\frac{1}{2\tau}\right|_{\textrm{TMA;}u\rightarrow\infty}=\frac{n_{i}}{\epsilon}\frac{4\pi^{2}v^{2}}{\pi^{2}+\mathcal{L}_{\text{II}}^{2}(\epsilon)},\,\,\eta_{zz}=0,\,\,\eta_{r}=0\,.\label{eq:tauUnitary}
\end{align}
In this case, considering $\lambda=10$ meV, $\Lambda=10\,\text{eV}$
and $n_{i},\epsilon$ as above one obtains a substantially shorter
scattering time $\left.\tau\right|_{\text{TMA}}=0.08$ ps. The unitary
result captures the typical energy dependence $\tau\propto\epsilon$
observed in high-mobility graphene samples \cite{Ferreira_11_res},
where the charge carrier mobility is likely limited by short-range
scatterers, including adsorbates, short-range ripples and vacancies
\cite{Chen_PRL2009,Monteverde_PRL2010,Ni_NanoLetters2010,Katoch_PRB2010}. 

\section{Microscopic linear response theory for spin relaxation}

\subsection{General formalism}

We consider the long-wavelength spin dynamics generated by a generic
external perturbation 
\begin{equation}
H_{\textrm{\ensuremath{\alpha\beta}}}^{\textrm{ext}}(\text{\textbf{x}},t)=-\mathcal{J}_{\alpha\beta}\mathcal{A_{\alpha\beta}}(\mathbf{x},t)\,,\label{eq:ext}
\end{equation}
where $\mathcal{J}_{\alpha\beta}\propto\sigma_{\alpha}s_{\beta}$
($\alpha,\beta=0,i$) is the current density operator ($\alpha=x,y$)
or density operator ($\alpha=0,z$) and $\mathcal{A_{\alpha\beta}}$
is a generalized vector potential \cite{Milletari_PRL2017}. We will
consider in detail two important cases: (i) an electric field perturbation
\emph{e.g.}, $H_{x0}^{\textrm{ext}}(\text{\textbf{x}},t)=-v\sigma_{x}s_{0}A_{x}(\mathbf{x},t)$
and (ii) a spin density fluctuation $H_{0i}^{\textrm{ext}}(\text{\textbf{x}},t)=-\frac{1}{2}\sigma_{0}s_{i}B_{i}(\mathbf{x},t)$.
The induced spin polarization density 
\begin{equation}
S_{i}(\mathbf{x},t)=\frac{1}{2}\langle\Psi^{\dagger}(\mathbf{x},t)\,\sigma_{0}s_{i}\,\Psi(\mathbf{x},t)\rangle\,,\label{eq:spin_polariz}
\end{equation}
is evaluated within the framework of linear response theory. This
approach has been applied to\textcolor{red}{{} }derive charge--spin
diffusion equations describing spin dynamics and magnetoelectric effects
in 2DEGs \cite{Maleki_CondMatt2017,Burkov_PRB2004,Burkov_PRB2004bis}.
As shown below, a suitable extension of this approach to accommodate
the enlarged (spin $\otimes$ pseudospin) Clifford algebra $\gamma_{\alpha\beta}=\sigma_{\alpha}s_{\beta}$
will allow us to obtain a rigorous microscopic theory of diffusive
transport and spin relaxation for 2D Dirac systems.

The linear response of the $\hat{i}$-component of the spin polarization
vector at zero temperature reads as 
\begin{equation}
S_{i}(\text{\textbf{x}},t)=-\int d\text{\textbf{x}}^{\prime}\int_{-\infty}^{\infty}dt^{\prime}\,\chi_{i,\alpha\beta}(\text{\textbf{x}}-\text{\textbf{x}}^{\prime},t-t^{\prime})\,\partial_{t^{\prime}}\mathcal{\mathcal{A_{\alpha\beta}}}(\text{\textbf{x}}^{\prime},t^{\prime})\,,\label{eq:ISGEconductivity}
\end{equation}
where $\chi_{i,\alpha\beta}(\text{\textbf{x}}-\text{\textbf{x}}^{\prime},t-t^{\prime})$
is the generalized spin susceptibility associated to the external
perturbation \emph{i.e.}, an electric field $\mathcal{E}_{x}(\mathbf{x},t)=-\partial_{t}A_{x}(\mathbf{x},t)$
or a 'spin injection field' $\Phi_{i}(\mathbf{x},t)=-\partial_{t}B_{i}(\mathbf{x},t)$
\cite{Shen_PRL2014}. Expressing the above equation in terms of the
Fourier transform $\chi_{i,\alpha\beta}(\mathbf{q},\omega)$ in the
long-wavelength limit $\text{\textbf{q}}\to0$ we have 
\begin{equation}
\chi_{i,\alpha\beta}(0,\omega)=\frac{\kappa}{2}\,\textrm{Tr}\left\langle \gamma_{0i}\,G^{R}(\mathbf{x},\mathbf{x}^{\prime};\epsilon+\omega)\,\gamma_{\alpha\beta}\,G^{A}(\mathbf{x}^{\prime},\mathbf{x};\epsilon)\right\rangle \,,\label{eq:spin_susceptibility}
\end{equation}
where $\kappa=v$ ($\kappa=1/2$) for a electric (spin injection)
field and Tr is the trace over all degrees of freedom. Terms involving
products of Green's functions on the same sector (RR and AA) are smaller
by a factor of $(\epsilon\tau)^{-1}$ and thus can be safely neglected.
\begin{figure}
\begin{centering}
\includegraphics[width=0.6\textwidth]{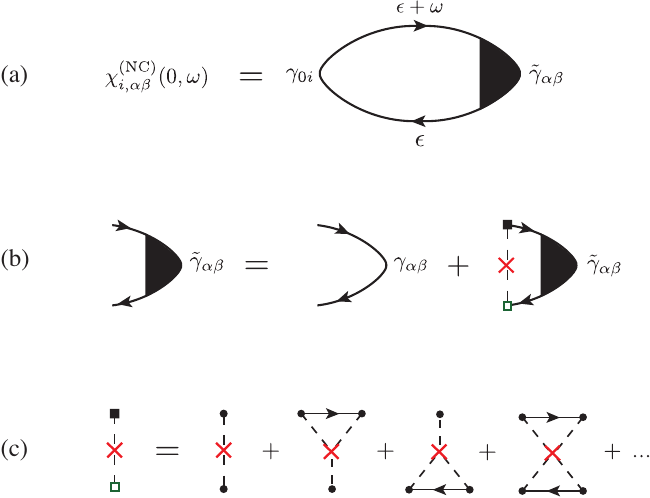} 
\par\end{centering}
\caption{\label{fig:02}Diagrammatic technique for evaluation of generalized
spin susceptibilities. (a) Two-particle ladder diagram. (b) BS equation
for the vertex renormalization. (c)\textcolor{blue}{{} }\textcolor{black}{Skeleton
expansion of the ladder diagram in terms of an infinite series of
two-particle, noncrossing diagrams. }Full (open) square denotes a
$T$($T^{\dagger}$) matrix insertion. }
\end{figure}

The disorder average in Eq.\,(\ref{eq:spin_susceptibility}) is evaluated
by means of the diagrammatic technique (Fig.\,\ref{fig:02}). For
brevity of notation, we first present the formalism within the Gaussian
approximation for the self-energy, Eq.\,(\ref{eq:SelfEnergyBA}).
In Sec.~\ref{subsec:SRTs_conservation_laws}, we provide the connection
with the full $T$-matrix result.

A summation of noncrossing two-particle (ladder) diagrams leads to
\begin{equation}
\chi_{i,\alpha\beta}^{(\textrm{NC})}(0,\omega)=\frac{\kappa}{2}\,\sum_{\mathbf{k}}\,\textrm{tr}\left\{ \gamma_{0i}\,\mathcal{G}_{\mathbf{k}}^{R}(\epsilon+\omega)\,\tilde{\gamma}_{\alpha\beta}(\omega)\,\mathcal{G}_{\mathbf{k}}^{A}(\epsilon)\right\} \,,\label{eq:spin_susceptibility-1}
\end{equation}
where tr is the trace over internal degrees of freedom (spin and sublattice).
The dressed vertex $\tilde{\gamma}_{\alpha\beta}$ satisfies the Bethe-Salpeter
(BS) equation 
\begin{equation}
\tilde{\gamma}_{\alpha\beta}(\omega)=\gamma_{\alpha\beta}+\frac{4}{2\pi\tau N_{0}}\,\sum_{\text{\textbf{k}}}\,\mathcal{G}_{\mathbf{k}}^{R}(\epsilon+\omega)\,\tilde{\gamma}_{\alpha\beta}(\omega)\,\mathcal{G}_{\mathbf{k}}^{A}(\epsilon)\,.\label{eq:BS}
\end{equation}
where $N_{0}\equiv\epsilon/\pi v^{2}$ (for the $T$-matrix extension
see Eq.~(\ref{eq:BS_TMatrix}) and text therein). Projecting onto
the elements of the Clifford algebra 
\begin{equation}
\tilde{\gamma}_{\alpha\beta\varrho\varsigma}(\omega)=\frac{1}{4}\text{tr}[\tilde{\gamma}_{\alpha\beta}(\omega)\,\sigma_{\varrho}s_{\varsigma}]\,,\label{eq:elements_CA}
\end{equation}
we recast the BS equation into the form 
\begin{equation}
\tilde{\gamma}_{\alpha\beta\varrho\varsigma}(\omega)=\delta_{\alpha\varrho}\delta_{\beta\varsigma}+\sum_{\mu,\nu=0,x,y,z}\,M_{\mu\nu\varrho\varsigma}(\omega)\tilde{\gamma}_{\alpha\beta\mu\nu}(\omega)\,,\label{eq:BSvertex}
\end{equation}
where 
\begin{equation}
M_{\mu\nu\varrho\varsigma}(\omega)=\frac{1}{2\pi\tau N_{0}}\sum_{\text{\textbf{k}}}\text{tr}\left[\mathcal{G}_{\mathbf{k}}^{R}(\epsilon+\omega)\gamma_{\mu\nu}\mathcal{G}_{\mathbf{k}}^{A}(\epsilon)\gamma_{\varrho\varsigma}\right]\,.
\end{equation}
Introducing the 16-dimensional vectors $\tilde{\boldsymbol{\gamma}}_{\alpha\beta}(\omega)=(\tilde{\gamma}_{\alpha\beta00}(\omega),...,\tilde{\gamma}_{\alpha\beta zz}(\omega))^{\text{t}}$
and $\boldsymbol{\gamma}_{\alpha\beta}=(0,0,....,\gamma_{\alpha\beta\alpha\beta},...,0)^{\textrm{t}}$
a more compact matrix form for Eq.\,(\ref{eq:BS}) is given in terms
of the \emph{diffuson operator $\mathcal{D}^{-1}$} as 
\begin{equation}
\mathcal{D}^{-1}\tilde{\mathbf{\boldsymbol{\gamma}}}_{\alpha\beta}(\omega)\equiv(\mathbf{1}_{16\times16}-M^{\textrm{t}}(\omega))\tilde{\mathbf{\boldsymbol{\gamma}}}_{\alpha\beta}(\omega)=\boldsymbol{\gamma}_{\alpha\beta}\,.\label{eq:diffuson}
\end{equation}
The spin relaxation rates are simply identified as the poles of the
generalized susceptibility in the complex $\omega$-plane. The determination
of the SRTs is thus reduced to the analysis of the behavior of $\mathcal{D}^{-1}=\mathcal{D}^{-1}(\omega)$
\cite{Rammer_Book1998}.

The formal result Eq.~(\ref{eq:diffuson}) deserves a few comments.
Firstly, $\mathcal{D}^{-1}$ spans in principle the entire Clifford
algebra, which physically encodes the coupled dynamics of spin and
other observables associated with the elements $\gamma_{\alpha\beta}$.
However, by exploiting symmetries, $\mathcal{D}^{-1}$ can be reduced
into block diagonal form, such that only some observables are coupled
to the spin polarizations along the three spatial directions. Secondly,
a distinct feature of Dirac systems is that spin densities are coupled
to charge currents even in the case (considered here) of spatially
uniform external perturbations $\text{\textbf{q}=0}$. The linear
Dirac dispersion of graphene is reflected in the form of the charge
current $J_{i}=v\sigma_{i}$ and spin current $\mathcal{J}_{i}^{a}=v\sigma_{i}s_{a}/2$
vertices, which do not depend explicitly on momentum; by virtue of
that they can be directly identified (apart from constants) as elements
of the Clifford algebra. Therefore all the relevant information about
coupling between currents and densities\emph{ is built-in} on the
$16\times16$ diffusion operator Eq.~(\ref{eq:diffuson}) in our
formalism. This will allows us to obtain a unified description of
spin relaxation processes and relativistic transport phenomena (e.g.,
charge-to-spin conversion) within our $\text{\textbf{q}=0}$ formalism\@.
We analyze the implications below.

The coupling of the electrons' spin to currents or other observables
in the long wavelength limit also suggests two equivalent scenarios
to study spin relaxation. The first natural choice is to consider
spin injection and investigate the relaxation of the spin density
profile (density--density response); alternatively, one can probe
the spin response indirectly by exciting an observable coupled to
the spin density through $\mathcal{D}^{-1}$. For instance, as we
will see in the following, one can drive a charge current via application
of an electric field to obtain a in-plane spin polarization of carriers
(ISGE). In that case, the information about the in-plane SRTs is readily
accessible by examining how the steady state (Edelstein) polarization
is achieved (density--current response). 

Before moving on, let us stress that within the Gaussian approximation,
a useful relation can be derived connecting the generalized susceptibility
Eq.\,(\ref{eq:spin_susceptibility-1}) and the renormalized vertex:
\begin{equation}
\chi_{i,\alpha\beta}^{\text{(NC)}}(0,\omega)=\frac{\kappa}{2\alpha}\sum_{\mu\nu}\,M_{\mu\nu0i}(\omega)\tilde{\gamma}_{\alpha\beta\mu\nu}(\omega)=\frac{\kappa}{2\alpha}\left(\tilde{\gamma}_{\alpha\beta0i}(\omega)-\delta_{\alpha0}\delta_{\beta i}\right)\,,\label{eq:ObservablesToVertex}
\end{equation}
where $\alpha\equiv(2\pi\tau N_{0})^{-1}$ and we have used Eq.\,(\ref{eq:BSvertex}).
The above equation states the spin response can be \emph{solely} obtained
from the associated component of the renormalized vertex $\tilde{\gamma}_{\alpha\beta0i}$.
A similar relation holds for other response functions. For example
the AC longitudinal (Drude) conductivity is written as 
\begin{equation}
\sigma_{xx}(\omega)=v^{2}\,\sum_{\mathbf{k}}\,\textrm{tr}\left\{ \gamma_{x0}\,\mathcal{G}_{\mathbf{k}}^{R}(\epsilon+\omega)\,\tilde{\gamma}_{x0}(\epsilon,\omega)\,\mathcal{G}_{\mathbf{k}}^{A}(\epsilon)\right\} =\frac{v^{2}}{\alpha}\left(\tilde{\gamma}_{x0x0}(\omega)-1\right)\,.\label{eq:ObservableToVertex_sigmaXX}
\end{equation}
Therefore Eq.\,(\ref{eq:ObservablesToVertex}) and similar relations
allow to identify the components of a renormalized vertex with the
associated observables, and will turn useful in the following.

Let us now determine the allowed couplings to $S_{x,y,z}$ by exploring
symmetry. The model of Eq.\,(\ref{eq:Hamiltonian}) is invariant
under the group $C_{\infty v}$, which is an emergent symmetry of
the continuum (long-wavelength) theory. As rotations in the continuum
do not describe the sublattice symmetry $A\leftrightarrow B$ of the
graphene system, a representation $U$ for the relevant set of discrete
operations has to be considered. Relevant to us are $C_{2}$, the
rotation of $\pi$ around the $\hat{z}$-axis exchanging sublattice
(and valleys), and $R_{x}$, the reflection over the $\hat{x}$-axis.
We have 
\begin{eqnarray}
U(C_{2}) & = & \tau_{x}s_{z}\,,\label{eq:symmetries}\\
U(R_{x}) & = & \tau_{z}\sigma_{x}s_{y}r_{x}.\label{eq:symmetries_2-1}
\end{eqnarray}
with $r_{x}:(\mathbf{x},\mathbf{y})\to(\mathbf{x},-\mathbf{y})$ and
$\tau_{i=x,y,z}$ are Pauli matrices acting on the valley degree of
freedom. We also make use of isospin (valley) rotations $\Lambda_{x,y,z}$
\cite{McCann06,Ostrovsky06} 
\begin{align}
\Lambda_{x,y} & =\tau_{x,y}\sigma_{z}\,,\label{eq:symmetries_2}\\
\Lambda_{z} & =\tau_{z}\,.\label{eq:symmetries_2_1}
\end{align}
For scalar disorder it suffices to examine the form of the clean-system
susceptibility at $\omega=0$ 
\begin{equation}
\mathcal{\chi}_{i,\alpha\beta}^{RA,\textrm{clean}}\equiv\frac{1}{4}\text{Tr}\ensuremath{\left[\gamma_{0i}\,G_{0}^{R}(\epsilon)\,\gamma_{\alpha\beta}\,G_{0}^{A}(\epsilon)\right]}.\label{eq:cleanSusceptibility}
\end{equation}
For any of the symmetries $\mathcal{S}$ listed above, we have $\mathcal{S}^{-1}G_{0}^{R/A}\mathcal{S}=G_{0}^{R/A}$,
and inserting resolutions of the identity in the form $\mathcal{S}^{\dagger}\mathcal{S}$
into Eq.\,(\ref{eq:cleanSusceptibility}) we find 
\begin{align}
\mathcal{\chi}_{i,\alpha\beta}^{RA,\textrm{clean}} & =\frac{p_{\alpha\beta}p_{0i}}{4}\text{Tr}\ensuremath{\left[\gamma_{0i}\,G_{0}^{R}(\epsilon)\gamma_{\alpha\beta}\,G_{0}^{A}(\epsilon)\right]}=p_{\alpha\beta}p_{0i}\,\mathcal{\chi}_{i,\alpha\beta}^{RA,\textrm{clean}}\,,
\end{align}
where $p_{\alpha\beta}(p_{0i})=\pm1$ is the parity of $\gamma_{\alpha\beta}(\gamma_{0i})$
under $\mathcal{S}$. From this result, we see that a nonzero response
requires the operator $\gamma_{\alpha\beta}$ to have the same parity
of the spin vertex under the action of any of $\mathcal{S}$. The
allowed couplings and parities under $\mathcal{S}$ are shown in the
Tab.\,\ref{tab:Table-Couplings}. As anticipated above, the in-plane
components $S_{x(y)}$ are coupled to orthogonal charge currents $\sigma_{y(x)}$,
as well as spin Hall currents $\gamma_{xz}(\gamma_{yz})$ and staggered
magnetizations $\gamma_{zy}(\gamma_{zx})$ \cite{Milletari_PRL2017,Offidani_PRL2017}.
The out-of-plane component $S_{z}$ is instead coupled to a mass term
$\sigma_{z}$ and in-plane spin currents $\gamma_{xx},\gamma_{yy}$. 
\begin{center}
\begin{table}[H]
\centering{}%
\begin{tabular}{|c|c|c|c|c|}
\hline 
Polarization  & $C_{2}$  & $R_{x}$  & $\Lambda_{x,y,z}$  & Couplings\tabularnewline
\hline 
\hline 
$S_{x}$  & -1  & -1  & +1  & $\sigma_{y},\,\gamma_{xz},\,\gamma_{zy}$\tabularnewline
\hline 
$S_{y}$  & -1  & +1  & +1  & $\sigma_{x},\,\gamma_{yz},\,\gamma_{zx}$\tabularnewline
\hline 
$S_{z}$  & +1  & -1  & +1  & $\sigma_{z},\,\gamma_{xx},\gamma_{yy}$\tabularnewline
\hline 
\end{tabular}\caption{\label{tab:Table-Couplings}Table summarizing the allowed couplings
to the spin polarizations in the 2D Dirac--Rashba model with nonmagnetic
scalar disorder.}
\end{table}
\par\end{center}

\subsection{Diffusive equations and SRTs}

In the following, we choose to consider the in-plane spin response
to an AC electric field $H_{\parallel}^{\text{ext}}=-v\sigma_{i}A_{i}(\omega)=-(i\omega)^{-1}v\sigma_{i}\mathcal{E}_{i}(\omega)$,
$i=x,y$. This choice, as discussed above, is equivalent to consider
in-plane spin injection, but has the advantage to allow for a unified
description of spin dynamics and charge-spin interconversion, e.g.
to capture the ISGE or other similar effects \cite{Huang_PRL2017,Note_1,Abanin_PRB2009}.
For the out-of-plane spin dynamics, we take a spin-density perturbation
$H_{\perp}^{\text{ext}}=\frac{1}{2}s_{z}B_{z}(\omega)$ (see Tab.
\ref{tab:Table-Couplings}).

\subsubsection{In-plane spin dynamics}

Without loss of generality, let us consider the dynamics of the $\hat{y}$
polarization. According to Tab.\,\ref{tab:Table-Couplings}, $s_{y}$
is coupled to three operators: $\sigma_{x},\,\sigma_{y}s_{z}$ and
$\sigma_{z}s_{x}$. However, leading terms in the $(\epsilon\tau)^{-1}$
expansion are only contained in the $s_{y}/\sigma_{x}$ sub-block.
Hence, to capture the SRTs it suffices to restrict to this $2\times2$
algebra. As anticipated above, we consider here the response to an
AC electric field $\mathcal{E}_{x}(\omega)$, associated with the
vertex $\kappa\gamma_{x0}=v\sigma_{x}\equiv v_{x}$. (Details of calculation
and full form of the $4\times4$ diffuson operator is given in Appendices
C and D.) To capture purely diffusive processes, we expand $\mathcal{D}^{-1}(\omega)$
in the low-frequency and small SOC limits, $\omega\tau\ll1$ and $\lambda\tau\ll1$,
respectively. In this regime, Eq.\,(\ref{eq:diffuson}) is written
then as 
\begin{equation}
\left(\begin{array}{cc}
\frac{1}{2}(1-\imath\omega\tau) & \frac{\lambda}{\epsilon}\Gamma_{s}(1+3\imath\omega\tau)\\
\frac{\lambda}{\epsilon}\Gamma_{\text{s}}(1+3\imath\omega\tau) & \Gamma_{s}-\imath\omega\tau
\end{array}\right)\left(\begin{array}{c}
\tilde{\gamma}_{x0x0}\\
\tilde{\gamma}_{x00y}
\end{array}\right)=\left(\begin{array}{c}
1\\
0
\end{array}\right)\,,\label{eq:PureDiffusiveM}
\end{equation}
where $\Gamma_{\text{s}}=\tau/\tau_{s}=2\lambda^{2}\tau$. In the
light of previous discussions (cf. Eqs.\,(\ref{eq:ObservablesToVertex})
and (\ref{eq:ObservableToVertex_sigmaXX})), $\tilde{v}_{x0}$ and
$\tilde{v}_{0y}$ are connected by a linear transformation to the
steady-state charge current and the spin polarization (Appendix D).

Off-diagonal elements of Eq.\,(\ref{eq:PureDiffusiveM}) carry in
relation to diagonal ones an extra order of smallness $\lambda/\epsilon$,
suggesting spin and charge to be weakly coupled in this limit. Their
inclusion however encodes charge-to-spin interconversion and it is
essential to get a correct physical description. The eigenvalues $-\imath\omega_{\pm}$
are found as 
\begin{align}
-\imath\omega_{+} & \simeq\frac{1}{\tau}\left(1+16\frac{\Gamma_{s}^{3}}{\epsilon^{2}\tau}\right)\,,\\
-\imath\omega_{-} & \simeq\frac{1}{\tau_{s}}\left(1-\frac{\Gamma_{s}^{3}}{\epsilon^{2}\tau}\right)\,,
\end{align}
and can be associated with charge current and spin relaxation times,
respectively. We see then the SRT can be identified as the mass ($\omega=0$)
term of the spin-spin part of the diffuson 
\begin{equation}
\frac{1}{\tau_{s}}\equiv\frac{1}{\tau_{s}^{\parallel}}\simeq1-M_{0y0y}(\omega=0)\simeq2\lambda^{2}\tau\,.\label{eq:SRTsimplified}
\end{equation}
Inverting Eq.\,(\ref{eq:PureDiffusiveM}), we find 
\begin{align}
\tilde{\gamma}_{x0x0} & \simeq\frac{1}{\tau}\frac{2}{-\imath\omega+\frac{1}{\tau}}\,,\\
\tilde{\gamma}_{x00y} & \simeq2\frac{\lambda}{\epsilon}\frac{1}{\tau}\frac{\Gamma_{s}}{-\imath\omega+\frac{\Gamma_{s}}{\tau}}\,,
\end{align}
from which, by using Eqs.\,(\ref{eq:ObservablesToVertex}) and (\ref{eq:ObservableToVertex_sigmaXX})
is it possible, upon Fourier transform, to derive the diffusive equation
of motion for coupled charge-spin dynamics as 
\begin{align}
\partial_{t}J_{x}(t) & =-\frac{1}{2\tau}(J_{x}(t)-J_{x}^{0}(t))\,,\label{eq:diffusion_eq1}\\
\partial_{t}S_{y}(t) & =-\frac{1}{\tau_{s}^{\parallel}}(S_{y}(t)-S_{y}^{0}(t))\,,\label{eq:diffusion_eq2}
\end{align}
where $J_{x}^{0}(t)\equiv2v^{2}\mathcal{E}_{x}(t)/\alpha$ and $S_{y}^{0}(t)\equiv-\lambda\mathcal{E}_{x}(t)/\epsilon\alpha$.
Note that charge current relaxation is regulated by the transport
time $\tau_{\text{tr}}\equiv2\tau$, indicating the absence of backscattering
\cite{Milletari_PRL2017,Milletari_Ferreira_QSHE_QTheory_16,Ferreira_11_res}.

\subsubsection{Out-of-plane spin dynamics}

For the out-of-plane spin dynamics we consider the renormalized vertex
$\kappa\tilde{\gamma}_{0z}=\frac{1}{2}\tilde{s}_{z}$. The off diagonal
components of the associated $4\times4$ diffuson block contains sub-leading
terms in the $(\epsilon\tau)^{-1}$ expansion (Appendix C), such that
the out-of-plane SRTs can be calculated similarly to Eq.\,(\ref{eq:SRTsimplified})
as 
\begin{equation}
\frac{1}{\tau_{s}^{\perp}}\simeq1-M_{0z0z}(\omega)\simeq4\lambda^{2}\tau\,.\label{eq:tauPerp}
\end{equation}
The generalization of the equations of motion Eqs.\,(\ref{eq:diffusion_eq1}),(\ref{eq:diffusion_eq2})
in this case is written as 
\begin{equation}
\partial_{t}S_{z}(t)=-\frac{1}{\tau_{s}^{\perp}}(S_{z}(t)-S_{z}^{0}(t))\,,
\end{equation}
where $S_{z}^{0}(t)=\dot{B}_{z}(t)/4\alpha$\textbf{ }is the effect
of the external perturbation (spin-injection field). The in-plane
and out-of-plane SRTs are in the following relation 
\begin{align}
\frac{1}{\tau_{s}^{\parallel}} & =\frac{1}{2}\frac{1}{\tau_{s}^{\perp}}\,,
\end{align}
which is nothing but the well-known DP ratio for 2DEGs \cite{Burkov_PRB2004}.
The above result has also been obtained for graphene within the time-dependent
perturbation theory for the density matrix \cite{Zhang_NJP2012}.
The agreement between graphene and the Rashba 2DEG results is expected
at high electronic density $\epsilon\gg\lambda$.

\subsection{SRT from the conservation laws in the DC limit\label{subsec:SRTs_conservation_laws}}

In this section, we demonstrate how the SRTs we have obtained above
can be equivalently extracted in the \emph{static limit $\omega=0$.
}This remarkable result is rooted in the conservation laws associated
to the disordered Dirac--Rashba Hamiltonian Eq.\,(\ref{eq:Hamiltonian})
\cite{Milletari_PRL2017}. The first step is to write the Heisenberg
equation of motion for the spin polarizations 
\begin{equation}
\partial_{t}S_{i}=\imath[H,S_{i}]=\frac{2\lambda}{v}\epsilon_{lj}\,\epsilon_{li}^{c}\,J_{j}^{c}\,,\label{eq:ConservationLaw}
\end{equation}
where $\epsilon_{lj},\epsilon_{li}^{c}$ are the second and third
rank Levi-Civita tensors and $J_{j}^{c}=\langle\mathcal{J}_{j}^{c}\rangle$
is the $\hat{j}$-component of the pure spin current (with polarization
$"c"$). As before, we consider an electric field applied along the
$\hat{x}$ direction. We find 
\begin{equation}
\partial_{t}S_{y}=\frac{2\lambda}{v}J_{y}^{z}\,,\label{eq:ConservationSY}
\end{equation}
where $J_{y}^{z}$ is identified as the spin Hall current according
to the chosen geometry. The spin Hall current is written in response
to the electric field 
\begin{equation}
J_{y}^{z}=\sigma_{yx}^{z}\mathcal{E}_{x}\,,
\end{equation}
where $\sigma_{yx}^{z}$ is the\textcolor{red}{{} }DC spin Hall conductivity
calculated according to Eq.\,\eqref{eq:spin_susceptibility-1} with
$\tilde{\gamma}_{0y}\to v\tilde{\gamma}_{yz}$. As for now no assumption
has been made for the self-energy approximation associated to the
scalar impurities field. Let us start from the more transparent Gaussian
case. Using the corresponding version of Eq.\,(\ref{eq:ObservablesToVertex})
for $\sigma_{yx}^{z}$, together with Eq.\,(\ref{eq:BSvertex}) we
have 
\begin{equation}
\sigma_{yx}^{z}=\frac{v^{2}}{2\alpha}\tilde{\gamma}_{x0yz}=\frac{v^{2}}{2\alpha}\left(M_{x0yz}\tilde{\gamma}_{x0x0}+M_{0yyz}\tilde{\gamma}_{x00y}\right)\,.\label{eq:SH1}
\end{equation}
In the latter we have neglected the terms $M_{yzyz}$ and $M_{zxyz}$
which, as said above, provide higher order corrections in the $(\epsilon\tau)^{-1}$
expansion. Multiplying both sides of Eq.\,(\ref{eq:SH1}) by the
electric field $\mathcal{E}_{x}$, and using $S_{y}=\chi_{y,x0}\,\mathcal{E}_{x}$
together with Eq.\,(\ref{eq:ObservablesToVertex}), we find 
\begin{equation}
J_{y}^{z}=\frac{v^{2}}{2\alpha}M_{x0yz}\tilde{\gamma}_{x0x0}\,\mathcal{E}_{x}+v\,M_{0yyz}S_{y}\,.\label{eq:JyZconservation}
\end{equation}
Despite the Dirac character of fermions, the steady-state case of
the continuity equation Eq.\,\ref{eq:ConservationLaw} imposes the
latter spin Hall current to vanish, analogously to the 2DEG case \cite{Milletari_PRL2017}.
This implies the establishment of the out-of-equilibrium value for
the spin polarization as
\begin{align}
S_{y}^{0}= & -\frac{\tilde{\gamma}_{x0x0}}{2\alpha v}\frac{M_{x0yz}}{M_{0yyz}}\mathcal{E}_{x}\,.\label{eq:EdelsteinSy0}
\end{align}
Evaluating the above quantities explicitly $\tilde{\gamma}_{x0x0}=2,\,M_{x0yz}/M_{0yyz}=\lambda/\epsilon$
and we recover the ISGE obtained in Ref.\,\cite{Offidani_PRL2017}.
Using Eq.\,(\ref{eq:ConservationSY}) we finally arrive at 
\begin{align}
J_{y}^{z} & =v\,M_{0yyz}\left(S_{y}-S_{y}^{0}\right)\label{eq:SHzero}
\end{align}
and therefore 
\begin{equation}
\partial_{t}S_{y}\equiv-\frac{1}{\tau_{s}^{\parallel}}\left(S_{y}-S_{y}^{0}\right)\,,\label{eq:syt}
\end{equation}
where we have identified the spin relaxation time 
\begin{equation}
\frac{1}{\tau_{s}^{\parallel}}=-2\lambda\,M_{0yyz}=2\lambda^{2}\tau\,,\label{eq:tauSparall_Bubble}
\end{equation}
in perfect accordance with the result obtained above, Eq.\,(\ref{eq:SRTsimplified}).
The bubble $M_{0yyz}$ is therefore what completely determines the
in-plane spin relaxation.

We now ask how the above result is modified when treating the self-energy
in the $T$-matrix approximation. The Bethe Salpeter equation Eq.\,(\ref{eq:BS})
now reads 
\begin{equation}
\tilde{\gamma}_{x0}(\epsilon)=\gamma_{x0}+n_{i}\sum_{\mathbf{k}}T^{R}(\epsilon)\,\mathcal{G}_{\mathbf{k}}^{R}(\epsilon)\,\tilde{\gamma}_{x0}(\epsilon)\,\mathcal{G}_{\mathbf{k}}^{A}(\epsilon)\,T^{A}(\epsilon)\,,\label{eq:BS_TMatrix}
\end{equation}
where $T^{R/A}(\epsilon)$ is the single-impurity $T$-matrix in the
R/A sectors introduced in Eq.\,(\ref{eq:TMatrix_SelfEnergy}). Projecting
onto the Clifford algebra, similarly to Eq.\,(\ref{eq:BSvertex}),
we have 
\begin{align}
\tilde{\gamma}_{x0\varrho\varsigma} & =\delta_{x\varrho}\delta_{0\varsigma}+\sum_{\mu\nu\zeta\xi=0,x,y,z}Y_{\varrho\varsigma\zeta\xi}N_{\mu\nu\zeta\xi}\tilde{\gamma}_{x0\mu\nu}\,,\label{eq:vertexTmatrix}
\end{align}
where we have defined 
\begin{align}
N_{\mu\nu\zeta\xi} & =\frac{n_{i}}{4}\sum_{\mathbf{k}}\text{tr}\,(\mathcal{G}_{\mathbf{k}}^{R}\,\gamma_{\mu\nu}\,\mathcal{G}_{\mathbf{k}}^{A}\,\gamma_{\zeta\xi})\,,\\
Y_{\varrho\varsigma\zeta\xi} & =\frac{1}{4}\text{tr}[T^{A}\,\gamma_{\varrho\varsigma}\,T^{R}\,\gamma_{\zeta\xi}]\,.
\end{align}
Recasting Eq.\,(\ref{eq:vertexTmatrix}) in vector notation, in the
same spirit of Eq.\,(\ref{eq:diffuson}), we have 
\begin{equation}
\tilde{\boldsymbol{\gamma}}_{x0}=\boldsymbol{\gamma}_{x0}+Y\,N^{\text{t}}\tilde{\boldsymbol{\gamma}}_{x0}\,,\label{eq:BS_Tmatrix}
\end{equation}
and consequently 
\begin{equation}
Y^{-1}(\tilde{\boldsymbol{\gamma}}_{x0}-\boldsymbol{\gamma}_{x0})=N^{\text{t}}\tilde{\boldsymbol{\gamma}}_{x0}\,.
\end{equation}
The latter equation allows again to find a connection with the observables.
For example, the generalization of Eq.\,(\ref{eq:ObservablesToVertex})
is written as 
\begin{equation}
\chi_{y,x0}=\frac{2v}{n_{i}}\sum_{\mu\nu}N_{\mu\nu0y}\tilde{\gamma}_{x0\mu\nu}=\frac{2v}{n_{i}}\sum_{\mu\nu}Y_{0y\mu\nu}^{-1}\tilde{\gamma}_{x0\mu\nu}\,.\label{eq:Susceptibility_Tmatrix}
\end{equation}
The spin Hall conductivity instead is found as 
\begin{equation}
\sigma_{yx}^{z}=\frac{2v^{2}}{n_{i}}\sum_{\mu\nu}Y_{yz\mu\nu}^{-1}\tilde{\gamma}_{x0\mu\nu}\,.\label{eq:SH_Tmatrix}
\end{equation}
Differently to the Gaussian case, where we were able to relate the
response of an observable uniquely to the associated component of
the renormalized vertex, in the $T$-matrix limit in principle all
components of $\tilde{\boldsymbol{\gamma}}_{x0}$ would contribute,
each of them with weight given by $Y^{-1}$. In the limiting case
of unitary limit $u_{0}\to\infty$, where $\lim_{u_{0}\to\infty}T^{R/A}=-\frac{1}{g_{0}^{R/A}}\,,$
we find a simplification as 
\begin{align}
Y_{\varrho\varsigma\zeta\xi}^{-1} & =|g_{0,0}^{R}|^{2}\delta_{\varrho\zeta}\delta_{\varsigma\xi}\,.
\end{align}
This implies that for Eq.(\ref{eq:SH_Tmatrix}) a relation similar
to the Gaussian case is obtained 
\begin{align}
J_{y}^{z} & =\sigma_{yx}^{z}\mathcal{E}_{x}=\frac{2v^{2}}{n_{i}}|g_{0,0}^{R}|^{2}\tilde{\gamma}_{x0yz}\,\mathcal{E}_{x}=\frac{2v^{2}}{n_{i}}N_{x0yz}\tilde{\gamma}_{x0x0}\mathcal{E}_{x}+v\,N_{00yz}|g_{0,0}^{R}|^{-2}S_{y}\,,
\end{align}
where we have restricted ourselves again to the dominant subspace
$\sigma_{x}/s_{y}$. After standard algebra, we arrive at 
\begin{align}
\partial_{t}S_{y} & =\frac{2\lambda}{v}\sigma_{yx}^{z}\mathcal{E}_{x}=\frac{2\lambda}{v}v\:N_{0yyz}(S_{y}-S_{y}^{0})\,,
\end{align}
and the SRT defined as 
\begin{equation}
\frac{1}{\tau_{s}^{\parallel}}=2\lambda\,|g_{0,0}^{R}|^{-2}N_{00yz}=2\lambda\frac{1}{\epsilon^{2}}\frac{16\pi^{2}v^{4}}{\pi^{2}+\mathcal{L}_{\text{II}}^{2}}N_{00yz}=2\lambda^{2}\tau\,,
\end{equation}
where we have used the definition of $\tau$ in the unitary limit,
Eq.\,(\ref{eq:tauUnitary}). We conclude that the the formal expression
connecting $\tau_{s}$ and $\tau$ (the DP relation) is the same as
found in the Gaussian limit for the self-energy. However, given the
different dependence of $\tau$ on the Fermi level in the two approximations---cf.
Eq.\,(\ref{eq:def_tau_0}) and Eq.\,(\ref{eq:tauUnitary})---one
has 
\begin{equation}
\frac{\tau(\epsilon)}{\tau_{s}^{\parallel}(\epsilon)}=\begin{cases}
\frac{2\lambda^{2}}{\epsilon^{2}}\left(\frac{2v^{2}}{n_{i}u_{0}^{2}}\right)^{2} & \text{Gaussian,}\\
\epsilon^{2}\frac{\lambda^{2}}{2}\left(\frac{\pi^{2}+\mathcal{L}_{\text{II}}^{2}}{4\pi^{2}n_{i}v^{2}}\right)^{2} & \text{Unitary}.
\end{cases}\label{eq:tauS_Gauss_Unitary}
\end{equation}
The SRT associated to the out-of-plane component can be derived along
the same lines. The relevant Heisenberg equation now reads 
\begin{equation}
\partial_{t}S_{z}=-\frac{2\lambda}{v}(J_{x}^{x}+J_{y}^{y})\,,
\end{equation}
and a similar reasoning that lead to Eq,\,(\ref{eq:tauSparall_Bubble}),
allows us to conclude 
\begin{equation}
\frac{1}{\tau_{s}^{\perp}}=2\lambda(M_{0zxx}+M_{0zyy})=4\lambda^{2}\tau\,,
\end{equation}
in the Gaussian limit, and a similar relation for the unitary limit. 

\subsection{Discussion\label{subsec:Discussion}}

\begin{figure}
\centering{}\includegraphics[width=0.4\columnwidth]{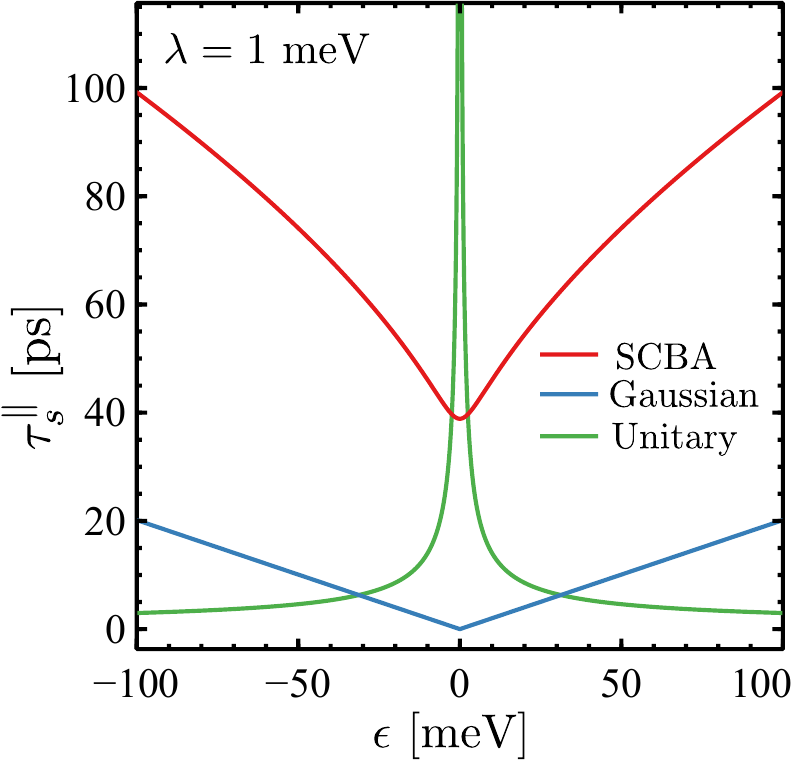}\caption{DP in-plane spin relaxation time calculated according different schemes
for the self-energy: SCBA (red line), Gaussian (blue line) and unitary
limit (green line). The most important feature obtained within the
SCBA is a strong renormalization of $\tau_{s}^{\parallel}$ in the
vicinity of the Dirac point, reflecting a disorder-induced finite
density of states in that region. In the plot $\lambda=1$ meV and
$\Gamma=60$ meV. \label{fig:DP-in-plane-spin}}
\end{figure}

Here, we discuss the DP relation obtained in Eq.\,\eqref{eq:tauS_Gauss_Unitary}
within the Gaussian and unitary limits of potential scattering. The
energy dependence of the spin lifetime for fixed impurity concentration
is shown in Fig.\,\ref{fig:DP-in-plane-spin}. Away from the Dirac
point, within the Gaussian approximation, the spin lifetime increases
linearly since $\tau\propto\epsilon^{-1}$ (see Eq.\,\eqref{eq:def_tau_0}).
In the unitary limit, instead, one has a linear dependence $\tau\propto\epsilon$
(see Eq.\,\eqref{eq:tauUnitary}), leading to vanishing spin lifetime
at high electron doping. On the other hand, near the Dirac point,
the noncrossing approximation breaks down. It is not surprising that
the spin lifetime dependences are found to be nonphysical as $\epsilon\to0$:
vanishingly small for the Gaussian limit and divergent for the unitary
limit. To overcome this limitation one needs to evaluate crossing
diagrams encoding quantum coherent processes, which includes weak
localization corrections and diffractive skew-scattering from two
or more impurities \cite{Milletari_Ferreira_QSHE_QTheory_16,Ado2015,Milletari_PRB(R)2016}.
However, here we are mostly interested in the diffusive regime away
from the Dirac point $\epsilon\tau\gg1$, thus neglecting interference
effects that can correct the standard DP relation \cite{Tuan_Sci2016,Cummings_PRL2016,McCann_PRL2012}.
However, an important refinement is possible within the noncrossing
formalism used here by evaluating the $O(n_{i}^{2}$) terms in Eq.~(\ref{eq:TMatrix_SelfEnergy}).
Such higher-order terms encode the strong renormalization of the single-particle
propagators by incoherent multiple scattering approaching the Dirac
point. To show this, it suffices to resum the infinite class of 'rainbow'
diagrams, a scheme known as self-consistent Born approximation (SCBA).
The SCBA self-energy is given by the solution of the following self-consistent
equation \cite{Ostrovsky06} 
\begin{equation}
\left.\frac{1}{2\tau}\right|_{\text{SCBA}}=-\Im\Sigma_{\text{SCBA}}(\epsilon)=-\Im\left[\frac{n_{i}}{4\pi v^{2}}(\epsilon-\Sigma_{\text{SCBA}}(\epsilon))\log\left(\frac{-\Lambda^{2}}{(\epsilon-\Sigma_{\text{SCBA}}(\epsilon))^{2}}\right)\right]\,.
\end{equation}

In Fig.\,\ref{fig:DP-in-plane-spin} we show that the SCBA provides
a physical (finite, nonzero) $\tau_{s}$ approaching the Dirac point.
To obtain a representative curve, we take $\lambda=1$~meV and we
choose the impurity density and the scattering strength such that
the SCBA nonperturbative energy scale $\Gamma=\Lambda\,e^{-2\pi v^{2}/(n_{i}u_{0}^{2})}$
\cite{Ostrovsky06} is a few tens meV. The in-plane SRT is then found
to lie in the range 50-100 ps. Concerning the magnitude of $\tau_{s}^{\parallel}$
we note the result is compatible with previous reports where the (uniform
or random) Rashba SOC is treated by semiclassical or numerical approaches
\cite{Tuan_Sci2016,Zhang_NJP2012}. 

\section{Conclusions}

In the present work, we laid the foundations of a general microscopic
theory of diffusive transport and spin relaxation in 2D Dirac systems
subject to spin--orbit interactions. Our work represents the logical
extension of the previously-developed diagrammatic treatments \cite{Burkov_PRB2004,Burkov_PRB2004bis}
to all orders in the scattering potential, for disordered electron
systems with an enlarged pseudospin $\otimes$ spin Clifford algebra
\cite{Milletari_PRL2017,Offidani_PRL2017}. We applied the formalism
to the paradigmatic case of 2D Dirac fermions with Rashba spin--orbit
coupling considering the purely diffusive regime $\lambda\tau\ll1\ll\epsilon\tau$.
We demonstrated how the Dyakonov-Perel relation between momentum and
spin lifetime $\tau\propto\tau_{s}^{-1}$ holds in both the Gaussian
(weak short-range scatterers) and the unitary (strong short-range
scatterers) limits, despite the drastic different dependence momentum
scattering times $\tau=\tau(\epsilon)$ in the two regimes. We derived
the same result both by direct diagrammatic resummation (in the noncrossing
approximation) and by exploiting the conservation laws of the theory
in the zero-frequency limit. Under the diffusive regime $\lambda\tau\ll1\ll\epsilon\tau$
is not possible to study the dynamics in the region of Fermi energies
comparable to the Rashba pseudogap region $\epsilon\sim2\lambda$,
which was recently predicted to display interesting out-of-equilibrium
phenomena \cite{Offidani_PRL2017}. The strong spin-momentum locking
approaching this regime lets us infer a modification of the relation
between $\tau_{s}$ and $\tau$ towards the Elliot-Yafet type $\tau_{s}\propto\tau$.
Our theory sets the stage to study the spin dynamics in that regime.
This topic has become of renewed great interest due to recent progresses
in graphene-based heterostructures, where the spin relaxation anisotropy
has been recognized as a viable tool to estimate the induced large
spin-orbital effects \cite{SRA_Cummings_17,SRA_Ghiasi_NanoLett2017,SRA_Benitez_2017,SRA_Wakamura_18}.

\section{Acknowledgements}

The authors are grateful to Ignacio Wilson-Rae for stimulating discussions.
M. O. and A. F. acknowledge funding from EPSRC (Grant No. EP/N004817/1).
Data availability statement (EPSRC).---No new data were created during
this study. 

\newpage{}

\appendix

\part*{Appendix}

\section{Clean Green's Function }

The explicit form of the clean single-particle Green's function is
\begin{equation}
\mathcal{G}_{0\mathbf{k}}^{a}(\epsilon)=-\frac{1}{2}\sum_{\mu=\pm1}\,L_{0\mu}^{a}\left[(\lambda+\mu\epsilon)\gamma_{0}+v\,\boldsymbol{\sigma}\cdot\mathbf{k}-\frac{\mu\epsilon}{2}\,\gamma_{r}+v\,(\mathbf{s}\times\mathbf{k})_{z}+\lambda\gamma_{zz}+\delta M_{2\phi_{\mathbf{k}}}\right]\,,\label{eq:G_0_explicit}
\end{equation}
where 
\begin{equation}
L_{0\mu}^{A(R)}=\frac{\mu}{v^{2}k^{2}-\epsilon^{2}-2\mu\lambda\epsilon\pm\imath\,0^{+}\text{sign}(\epsilon-\mu\lambda)}\,,\label{eq:kernel}
\end{equation}
\begin{equation}
\delta M_{2\phi_{\mathbf{k}}}=-\frac{1}{2}(\epsilon+2\mu\lambda)\left[(\sigma_{y}s_{y}-\sigma_{x}s_{x})\sin2\phi_{\mathbf{k}}+(\sigma_{x}s_{y}+\sigma_{y}s_{x})\cos2\phi_{\mathbf{k}}\right],\label{eq:G_2phi}
\end{equation}
and $\phi_{\mathbf{k}}$ is the angle formed by the wavevector with
$\hat{k}_{x}$.

\section{Integrals and expansion}

The current work makes extensive use of momentum integrations involving
products of two renormalized Green's functions with analyticity in
opposite halves of the complex plane, see e.g. Eqs.\,(\ref{eq:spin_susceptibility}),(\ref{eq:BS}).
The retarded function is displaced in energy by the amount $\omega$.
Similarly to the bare Green's function decomposition Eqs.\,(\ref{eq:G_0_explicit})
and (\ref{eq:G_2phi}), we write the renormalized (disorder averaged)
propagators as 
\begin{equation}
\mathcal{G}_{\text{\textbf{k}}}^{a}(\epsilon)=M_{1\text{\textbf{k}}}^{a}(\epsilon)\,L_{1\boldsymbol{\text{\textbf{k}}}}^{a}(\epsilon)+M_{2\text{\textbf{k}}}^{a}(\epsilon)\,L_{2\text{\textbf{k}}}^{a}(\epsilon)\,,\label{eq:G_ren_decomposed}
\end{equation}
where $M_{i\text{\textbf{k}}}^{a}(\epsilon)=M_{i}^{a\,(0)}+v^{2}k^{2}M_{i}^{a\,(2)},\,\,i=\{1,2\}$
are matrix coefficients and the kernels $L_{\mu}^{a}=L_{i\text{\textbf{k}}}^{a}$
are obtained in the Gaussian limit from the functions $L_{0\mu}$
of Eq.\,(\ref{eq:kernel}) by analytical continuation $\epsilon\to\epsilon+a\frac{\imath}{2\tau}$.
In the $T$ matrix approach, the analytical continuation has to be
performed as to include the other matrix structure of the self-energy
$\propto\gamma_{r},\gamma_{\text{KM}}$ \cite{Offidani_PRL2017}.
We can generically write 
\begin{equation}
L_{i\text{\textbf{k}}}^{a}(\epsilon)=\frac{1}{v^{2}k^{2}-z_{i}^{a}(\epsilon)}\,,
\end{equation}
where $z_{i}^{a}(\epsilon)$ are complex quantities. Given the decomposition
in Eq.\,(\ref{eq:G_ren_decomposed}), the integrals we need to solve
are reduced to product of two kernels in different combinations, accompanied
or not by a factor $v^{2}k^{2}$. Terms proportional to $v^{4}k^{4}$
can be shown to vanish upon angular integration $\int d\phi_{\text{\textbf{k}}}$.
We write below an \emph{exact }solution and then expand at linear
order in $\omega$. For simplicity we show the results for the Gaussian
approximation. The first type of integrals is 
\begin{align}
\Gamma_{ij}=\int_{0}^{\infty}\frac{dk\,k}{2\pi}\,L_{i\text{\textbf{k}}}^{R}(\epsilon+\omega)\,L_{j\text{\textbf{k}}}^{A}(\epsilon) & =\int_{0}^{\infty}\frac{dk\,k}{2\pi}\,\frac{1}{v^{2}k^{2}-z_{i\text{\textbf{k}}}^{R}(\epsilon+\omega)}\frac{1}{v^{2}k^{2}-z_{j\text{\textbf{k}}}^{A}(\epsilon)}\\
= & \frac{1}{z_{i\text{\textbf{k}}}^{R}(\epsilon+\omega)-z_{j\text{\textbf{k}}}^{A}(\epsilon)}\int_{0}^{\infty}\frac{dk\,k}{2\pi}\left(\frac{1}{v^{2}k^{2}-z_{i\text{\textbf{k}}}^{R}(\epsilon+\omega)}-\frac{1}{v^{2}k^{2}-z_{j\text{\textbf{k}}}^{A}(\epsilon)}\right)\\
= & \frac{1}{4\pi v^{2}}\frac{1}{z_{i\text{\textbf{k}}}^{R}(\epsilon+\omega)-z_{j\text{\textbf{k}}}^{A}(\epsilon)}\times\\
 & \left[-\log\left(-z_{i\text{\textbf{k}}}^{R}(\epsilon+\omega)\right)+\log\left(-z_{j\text{\textbf{k}}}^{A}(\epsilon)\right)-\left(\frac{1}{z_{i\text{\textbf{k}}}^{R}(\epsilon)}\partial_{\epsilon}z_{i\text{\textbf{k}}}^{R}(\epsilon)\right)\omega\right]\,,
\end{align}
where the principal branch of the log function has been chosen. Note
$z_{1}^{a}(\lambda)\to z_{2}^{a}(-\lambda)$. Thus, $\Gamma_{11}(\lambda)=\Gamma_{22}(-\lambda)$
and $\Gamma_{12}(\lambda)=\Gamma_{21}(-\lambda)$. At linear order
in $\omega$ we find 
\begin{align}
\Gamma_{11} & =\frac{1}{4\pi v^{2}}\left[\frac{\pi}{\epsilon+\lambda}-\frac{1}{\epsilon(\epsilon+2\lambda)}+\imath\pi\omega\tau\frac{\tau}{\epsilon+\lambda}\right],\\
\Gamma_{22} & =\Gamma_{11}(\lambda\to-\lambda)\,,\\
\Gamma_{12}= & \frac{1}{4\pi v^{2}}\left[\frac{2\imath\pi}{4\epsilon\lambda}-\omega\frac{\epsilon+\lambda}{2\epsilon^{2}\lambda(\epsilon+2\lambda)}\left(1+\imath\pi\frac{\epsilon+2\lambda}{2\lambda}\right)\right]\,,\\
\Gamma_{21} & =\Gamma_{12}(\lambda\to-\lambda)\,,
\end{align}
where we have retained leading order terms in $(\epsilon\tau)^{-1}$.
The other class of integrals we need to solve is 
\begin{align}
\Gamma_{ij}^{(3)} & =\int_{0}^{\Lambda/v}\frac{dk\,k^{3}}{2\pi}L_{i\text{\textbf{k}}}^{R}(\epsilon+\omega)\,L_{j\text{\textbf{k}}}^{A}(\epsilon)=\int_{0}^{\Lambda/v}\frac{dk\,k^{3}}{2\pi}\,\frac{1}{v^{2}k^{2}-z_{i\text{\textbf{k}}}^{R}(\epsilon+\omega)}\frac{1}{v^{2}k^{2}-z_{j\text{\textbf{k}}}^{A}(\epsilon)}\\
 & =\frac{1}{4\pi v^{2}}\frac{}{z_{i\text{\textbf{k}}}^{R}(\epsilon+\omega)-z_{j\text{\textbf{k}}}^{A}(\epsilon)}\left[2\imath\,\text{Im}\left(z_{i\text{\textbf{k}}}^{R}(\epsilon)\log\left(\frac{\Lambda^{2}}{-z_{i\text{\textbf{k}}}^{R}(\epsilon)}\right)\right)-\partial_{\epsilon}z_{i\text{\textbf{k}}}^{R}(\epsilon)\left(1-\log\left(\frac{\Lambda^{2}}{-z_{i\text{\textbf{k}}}^{R}(\epsilon)}\right)\right)\omega\right]\,,
\end{align}
where the ultraviolet cutoff $\Lambda/v\gg k_{\text{F}}$ has been
introduced to regularize the integrals. A careful evaluation yields
\begin{align}
\Gamma_{11}^{(3)} & =\frac{1}{4\pi v^{2}}\left[\frac{\pi\epsilon(\epsilon+2\lambda)\tau}{\epsilon+\lambda}(1+\imath\omega\tau)+\log\left|\frac{\Lambda^{2}}{\epsilon^{2}+2\epsilon\lambda}\right|-1-\omega\frac{\pi\epsilon\lambda\tau}{2(\epsilon+\lambda)^{2}}\right]\,,\\
\Gamma_{12}^{(3)} & =\frac{1}{4\pi v^{2}}\left[\frac{2\imath\pi\epsilon+2\lambda\,\mathcal{L}_{\text{II}}}{4\lambda}-\omega\frac{\epsilon+\lambda}{2\epsilon\lambda}\left(1+2\pi\frac{\epsilon+2\lambda}{4\lambda}\right)\right]\,,
\end{align}
and the expressions for $1\leftrightarrow2$ again obtainable with
the replacement $\lambda\to-\lambda.$

\section{Full form of the diffuson}

Here we report the full form of the two relevant blocks of the diffuson,
involving $S_{y,z}$. The expressions are provided at leading order
in the expansions for $\omega\tau\ll\lambda\tau\ll1\ll\epsilon\tau$. 
\begin{itemize}
\item Subspace $\sigma_{x},\,s_{y},\,\sigma_{y}s_{z},\,\sigma_{z}s_{x}$
\begin{align}
\left.\mathcal{D}^{-1}\right|_{s_{y}} & =\left(\begin{array}{cccc}
\frac{1}{2}(1-\imath\omega\tau) & \frac{2\lambda^{3}\tau^{2}}{\epsilon}(1+3\imath\omega\tau) & -\frac{\lambda^{2}\tau}{\epsilon}(1+2\imath\omega\tau) & \frac{\lambda^{3}\tau}{\epsilon^{2}}(1+2\imath\omega\tau)\\
\\
\frac{2\lambda^{3}\tau^{2}}{\epsilon}(1+3\imath\omega\tau) & 2\lambda^{2}\tau^{2}-\imath\omega\tau & -\lambda\tau(1+2\imath\omega\tau) & \frac{\lambda^{2}\tau}{\epsilon}(1+2\imath\omega\tau)\\
\\
\frac{\lambda^{2}\tau}{\epsilon}(1+2\imath\omega\tau) & \lambda\tau(1+2\imath\omega\tau) & \frac{1}{2}(1-\imath\omega\tau) & \frac{\lambda}{2\epsilon}(1+\imath\omega\tau)\\
\\
-\frac{\lambda^{3}\tau}{\epsilon^{2}}(1+2\imath\omega\tau) & -\frac{\lambda^{2}\tau}{\epsilon}(1+2\imath\omega\tau) & \frac{\lambda}{2\epsilon}(1+\imath\omega\tau) & 1-\imath\omega\tau\,\frac{\lambda^{2}}{2\epsilon^{2}}
\end{array}\right)\,,
\end{align}
\item Subspace $s_{z},\,\sigma_{x}s_{x},\,\sigma_{y}s_{y},\,\sigma_{z}$
\begin{align}
\left.\mathcal{D}^{-1}\right|_{s_{z}} & =\left(\begin{array}{cccc}
4\lambda^{2}\tau^{2}-\imath\omega\tau & \lambda\tau(1+2\imath\omega\tau) & \lambda\tau(1+2\imath\omega\tau) & -\frac{\lambda}{\pi\tau\epsilon^{2}}+O[(\epsilon\tau)^{-4}]\\
\\
-\lambda\tau(1+2\imath\omega\tau) & \frac{1}{2}(1-\imath\omega\tau) & \frac{\lambda^{2}\tau^{2}}{2}(1+3\imath\omega\tau) & O[(\epsilon\tau)^{-4}]\\
\\
-\lambda\tau(1+2\imath\omega\tau) & \frac{\lambda^{2}\tau^{2}}{2}(1+3\imath\omega\tau) & \frac{1}{2}(1-\imath\omega\tau) & O[(\epsilon\tau)^{-4}]\\
\\
-\frac{\lambda}{\pi\tau\epsilon^{2}}+O[(\epsilon\tau)^{-4}] & O[(\epsilon\tau)^{-4}] & O[(\epsilon\tau)^{-4}] & 1+\frac{\imath\omega}{4\epsilon^{2}\tau}
\end{array}\right)\,.
\end{align}
\end{itemize}

\section{Equation for operators instead of vertices}

In the main text, we have written equations of motion for the renormalized
vertices, rather than for the observables themselves. As an example,
here we report the diffusive matrix $\mathcal{D}^{-1}$ for the observables,
in the relevant sub-block $s_{y}/\sigma_{x}$ for the in-plane spin
dynamics. Also here we consider the response to an external electric
field $\mathcal{E}_{x}$. To this aim we recall in the Gaussian approximation
(cf. Eq.\,(\ref{eq:ObservablesToVertex}) and (\ref{eq:ObservableToVertex_sigmaXX}))
\begin{align}
J_{x} & =\sigma_{xx}\,\mathcal{E}_{x}=\frac{v^{2}}{\alpha}(\tilde{\gamma}_{x0x0}-1)\mathcal{E}_{x}\,,\\
S_{y} & =v\,\chi_{y,0x}\,\mathcal{E}_{x}=\frac{v}{2\alpha}\tilde{\gamma}_{x00y}\mathcal{E}_{x}\,.
\end{align}
Manipulating Eq.\,(\ref{eq:diffuson}) we have 
\begin{align}
\mathcal{D}^{-1}\boldsymbol{\tilde{\gamma}}_{x0}=\boldsymbol{\gamma}_{x0} & \implies\mathcal{C}\boldsymbol{\tilde{\gamma}}_{x0}=\mathcal{C}\mathcal{D}\boldsymbol{\gamma}_{x0}\,,
\end{align}
where we have defined the matrix 
\begin{equation}
\mathcal{C}=\frac{v\,\mathcal{E}_{x}}{\alpha}\text{diag}(v,\frac{1}{2})\,.
\end{equation}
Consequently by subtracting to both sides $v^{2}\mathcal{E}_{x}\boldsymbol{\gamma}_{x0}/\alpha$
we have 
\begin{equation}
\frac{v\,\mathcal{E}_{x}}{\alpha}\left[\left(\begin{array}{c}
v\,\tilde{\gamma}_{x0x0}\\
\frac{\tilde{\gamma}_{x0x0}}{2}
\end{array}\right)-\left(\begin{array}{c}
v\\
0
\end{array}\right)\right]\equiv\left(\begin{array}{c}
J_{x}\\
S_{y}
\end{array}\right)=(\mathcal{CD}-\frac{v^{2}}{\alpha}\mathcal{E}_{x}\mathds{1})\boldsymbol{\gamma}_{x0}\,.
\end{equation}
We conclude the diffusive matrix for the observables is 
\begin{equation}
\mathcal{D}_{\text{obs}}^{-1}=(\mathcal{CD}-\frac{v^{2}}{\alpha}\mathcal{E}_{x}\mathds{1})^{-1}\,.
\end{equation}
Direct inspection shows that $\mathcal{D}_{\text{obs}}^{-1}$ and
$\mathcal{D}^{-1}$ share the same pole structure, justifying the
approach in the main text.

\end{document}